\documentclass[11pt]{article}
\usepackage{graphicx}
\usepackage{amsmath,amssymb}
\def\1{{\bf 1}}

\def\be{\begin{equation}}
\def\ee{\end{equation}}

\begin{document}
\noindent {\large \bf The Structural Stability of Wild-type Horse Prion Protein}\\

\bigskip

\noindent {\Large Jiapu Zhang}\\
School of Sciences, Information Technology and Engineering,\\ 
University of Ballarat, Mount Helen, Ballarat, Victoria 3353, Australia,\\
Phone: 61-423487360, Email: jiapu\_zhang@hotmail.com, j.zhang@ballarat.edu.au\\ 

\bigskip

\noindent {\bf \Large Abstract}\\

\noindent Prion diseases {\it (e.g. Creutzfeldt-Jakob disease (CJD), variant CJD (vCJD), Gerstmann-Str$\ddot{\text{a}}$ussler-Scheinker syndrome (GSS), Fatal Familial Insomnia (FFI) and Kuru in humans, scrapie in sheep, bovine spongiform encephalopathy (BSE or `mad-cow' disease) and chronic wasting disease (CWD) in cattles)} are invariably fatal and highly infectious neurodegenerative diseases affecting humans and animals. However, by now there have not been some effective therapeutic approaches or medications to treat all these prion diseases. Rabbits, dogs, and horses are the only mammalian species reported to be resistant to infection from prion diseases isolated from other species. Recently, the $\beta$2-$\alpha$2 loop has been reported to contribute to their protein structural stabilities. The author has found that rabbit prion protein has a strong salt bridge ASP177-ARG163 (like a taut bow string) keeping this loop linked. This paper confirms that this salt bridge also contributes to the structural stability of horse prion protein. Thus, the region of $\beta$2-$\alpha$2 loop might be a potential drug target region. Besides this very important salt bridge, other four important salt bridges GLU196-ARG156-HIS187, ARG156-ASP202 and GLU211-HIS177 are also found to greatly contribute to the structural stability of horse prion protein. Rich databases of salt bridges, hydrogen bonds and hydrophobic contacts for horse prion protein can be found in this paper.\\


\noindent {\bf Abbreviations:} CJD: Creutzfeldt-Jakob disease, vCJD: variant Creutzfeldt-Jakob diseases, GSS: Gerstmann-Straussler-Scheinker syndrome, FFI: Fatal Familial Insomnia, BSE: bovine spongiform encephalopathy, CWD: chronic wasting disease, MD: molecular dynamics, RMSD: root mean square deviation, HoPrP$^{\text{C}}$: horse prion protein, DoPrP$^{\text{C}}$: dog prion protein, RaPrP$^{\text{C}}$:  rabbit prion protein.\\

\section*{Introduction}
Prion diseases such as CJD, vCJD, GSS, FFI, Kuru in humans, scrapie in sheep, BSE or `mad-cow' disease and CWD in cattle are invariably fatal and highly infectious neurodegenerative diseases affecting humans and animals. However, by now there have not been some effective therapeutic approaches or medications to treat all these diseases (1-3). In 2008, canine mammals including dogs (canis familials) were the first time academically reported to be resistant to prion diseases (4) (where, before 2008, dogs and horses were reported to resist prion diseases in media). Rabbits are the mammalian species known to be resistant to infection from prion diseases from other species (5). Horses were academically reported to be resistant to prion diseases too (6). Thus, it is very worth studying the molecular structures of dog, rabbit and horse prion proteins to obtain insights into the immunity of dogs, rabbits and horses to prion diseases. Molecular dynamics (MD) simulations provide an excellent method to understand the structural stability as well as ligand interactions of biological systems (7-38). It should be noted that recently there have been a devoted effort by MD to understand the human prion proteins (39-42). This author has already investigated (43-45) by MD simulations the dog and rabbit prion proteins. This paper is focusing on the MD simulation studies on the horse prion protein C-terminal structured region, as well as a comparison of rabbit, dog and horse prion proteins.\\

Rabbits, dogs and horses are resistant to prion diseases. The infectious prion (PrP$^{\text{Sc}}$) is an abnormally folded form of the normal cellular prion (PrP$^{\text{C}}$) and the conversion of PrP$^{\text{C}}$ to PrP$^{\text{Sc}}$ is believed to involve conformational change from a predominantly $\alpha$-helical protein (42\% $\alpha$-helix, 3\% $\beta$-sheet) to a protein rich in $\beta$-sheets (30\% $\alpha$-helix, 43\% $\beta$-sheet). For a protein structure, its stability is maintained by its hydrogen bonds, hydrophobic bonds, salt bridges, and van der Waals contacts. Thus, when we study their NMR structure and structural dynamics, we have to consider these factors. The secondary structure of rabbit prion protein turned into $\beta$-sheet structures from $\alpha$-helical structures under low pH environment. The change of pH environments causes the change of secondary structure from a-helices into ß-sheets, because the salt bridges disappear under low pH environment. Thus, we might say that the salt bridge provides the stability and it might be be a drug target.\\

All the MD simulations in this paper confirmed the structural stability of wild-type horse prion protein under both neutral and low pH environments. The analyses of salt bridges, hydrogen bonds and hydrophobic contacts for horse prion protein will be done in order to seek reasons of the stability (where the salt bridges, hydrogen bonds and hydrophobic contacts will be presented in this paper at the residue-residue level). The rest of this paper is arranged as follows. We introduce the MD simulation materials and methods (similar as (44-46)) in Section 2. Section 3 mainly gives MD simulation results and their discussions. Concluding remarks are given in the last section.\\

\section*{Materials and Methods}
The MD simulation materials and methods for horse prion protein are completely same as the ones of (44-45). Simulation initial structure for the horse prion protein was built on HoPrP$^{\text{C}}$(119-231) (PDB entry 2KU4). The low pH in the simulations is achieved by the change of residues HIS, ASP, GLU into HIP, ASH, GLH respectively and the Cl- ions added by the XLEaP module of AMBER 11. The neutral pH in the simulations is achieved by the change of residues HIS into HID and the Na+ ions added by the XLEaP module of AMBER 11. 16 Cl- and 6281 waters were added for the horse prion protein under low pH environment, and 2 Na+ and 6679 waters were added under neutral pH environment. After equilibrations, 30 ns' production MD simulations were done using constant pressure and 350 K temperature ensemble for the horse prion protein for both the {\it seed1} and {\it seed2} defined in (45). The two seeds are two different initial velocities. Different initial conditions should produce the same thermodynamic quantities after equilibration. This will firmly ensure that our research findings in this paper are correct.\\    

The studies on the rabbit prion protein at 300 K, 450 K, 500 K (43-46) and the dog prion protein at 300 K, 450 K (45) have confirmed the research findings on the salt bridge ASP178-ARG164 and the secondary structure change of rabbit prion protein from neutral to low pH environments. 350 K is a practical experimental laboratory temperature for prion proteins reported. Thus, 350 K is set for HoPrP$^{\text{C}}$ in this paper.\\ 

\section*{Results and Discussions}
\begin{figure}[h!]
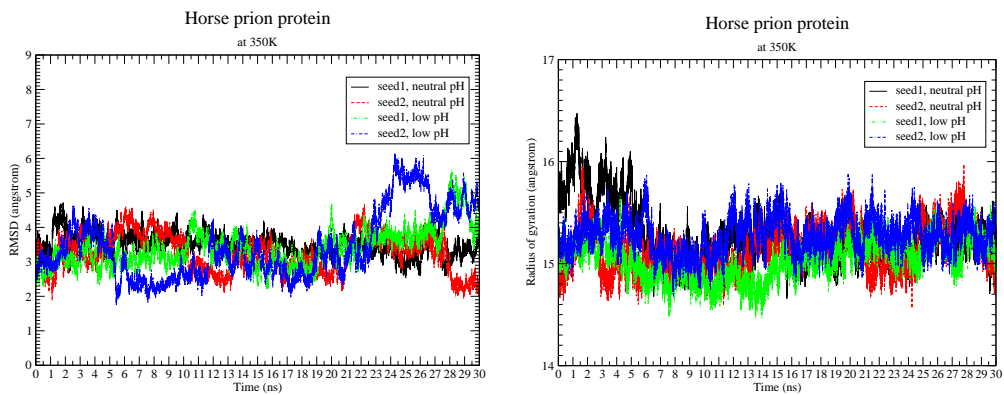

\centerline{
\includegraphics[width=2.5in]{RMSD_horse_350K_seed1and2_productions_01to30ns.eps}\\
\includegraphics[width=2.5in]{Radius_horse_350K_seed1and2_production01to30ns.eps}
}
\caption{The RMSD and Radius Of Gyration graphs for horse prion protein at 350K.}
\label{RMSD_Radius}
\end{figure}

\begin{figure}[h!]
\centerline{
\includegraphics[width=5.5in]{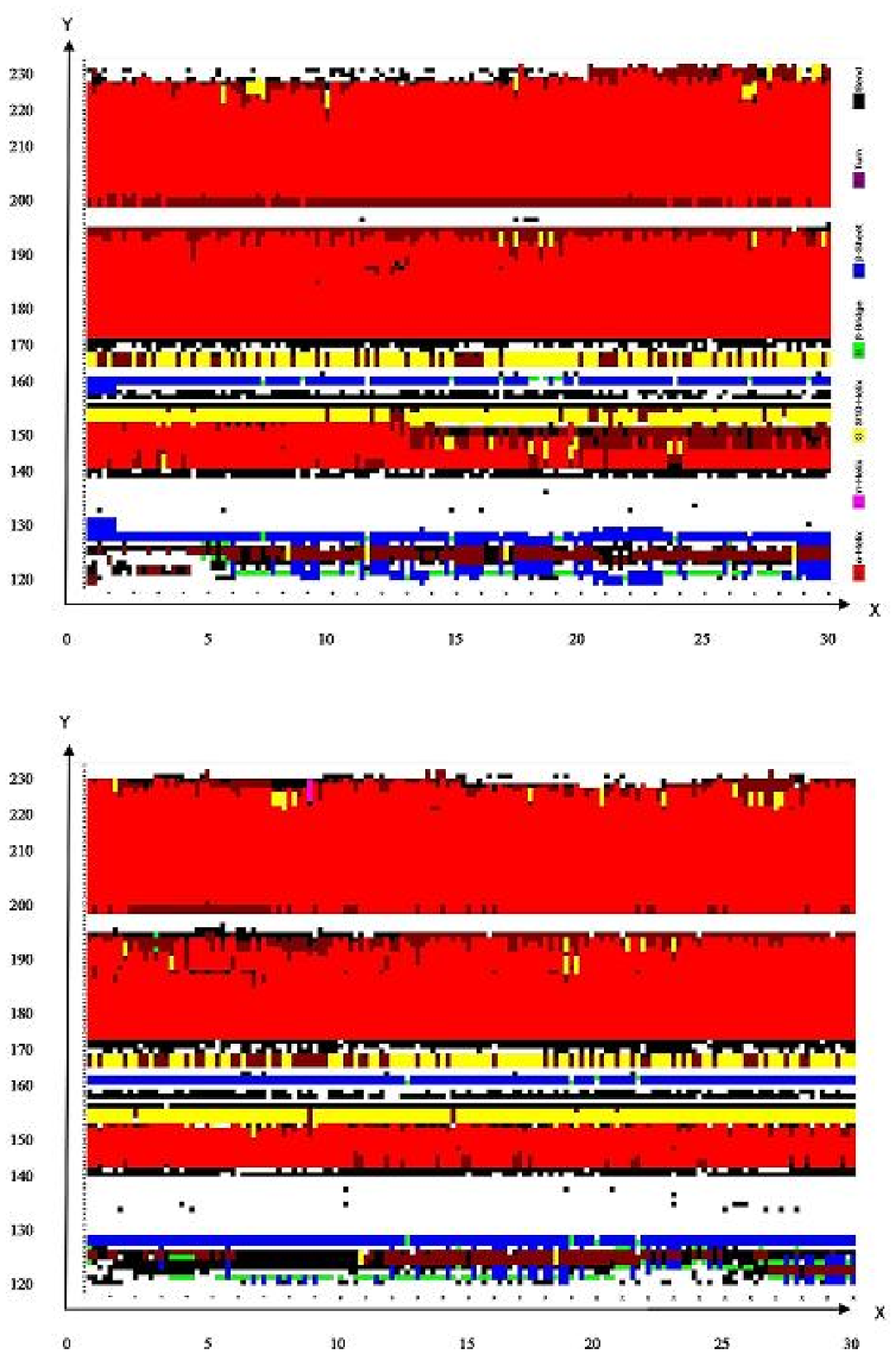}
}
\caption{Secondary structures of horse prion protein at 350 K under \underline{neutral} pH environment: {\it seed1} and {\it seed2} (from up to down). (X-axis: time (from left to right: 0 ns - 30 ns), Y-axis: residue numbers (from down to up: 119 - 231).)}
\label{secondary_structures1}
\end{figure}

\begin{figure}[h!]
\centerline{
\includegraphics[width=5.5in]{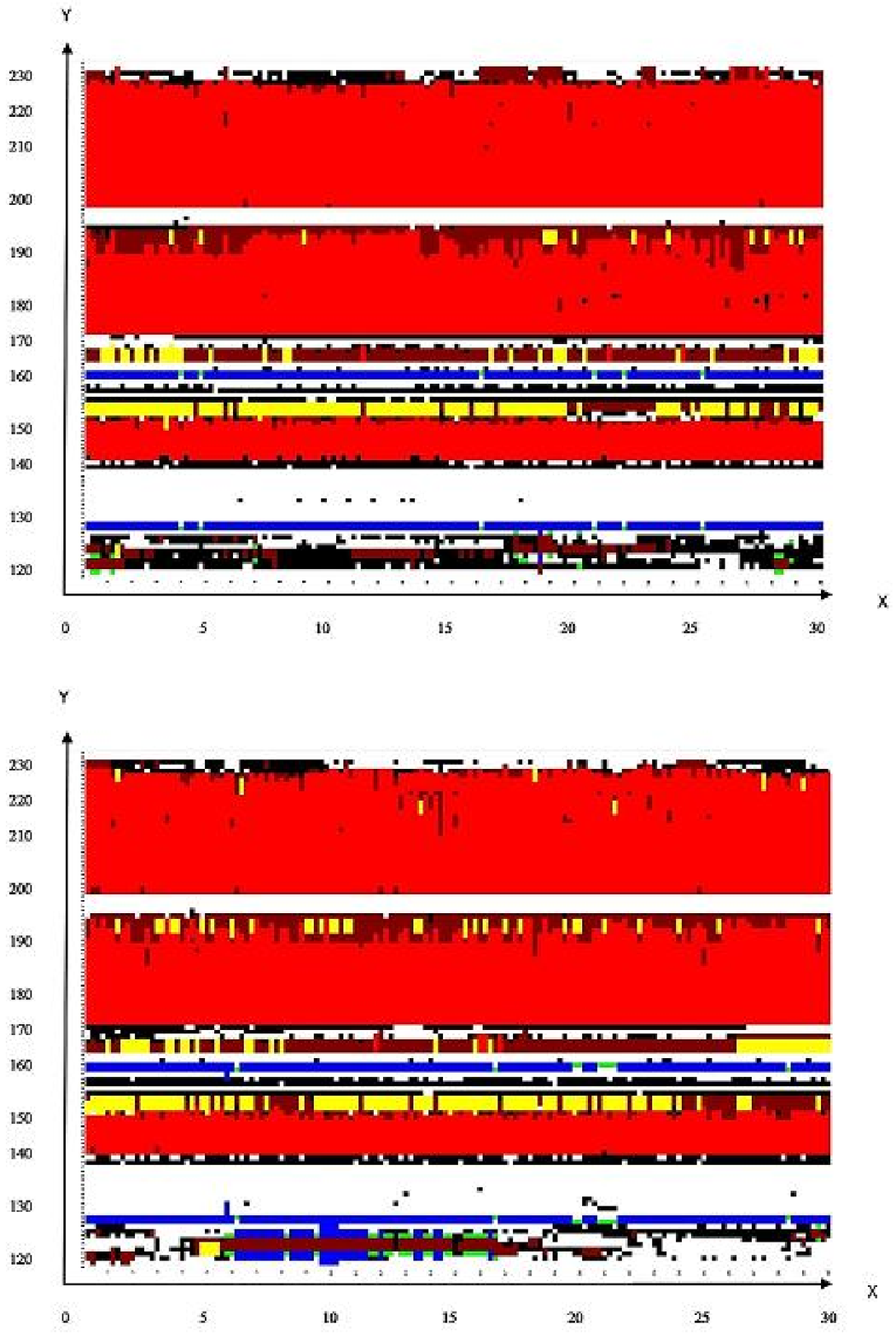}
}
\caption{Secondary structures of horse prion protein at 350 K under \underline{low} pH environment: {\it seed1} and {\it seed2} (from up to down). (X-axis: time (from left to right: 0 ns - 30 ns), Y-axis: residue numbers (from down to up: 119 - 231).)}
\label{secondary_structures2}
\end{figure}

For both the {\it seed1} and {\it seed2}, the MD simulations at room temperature 300 K whether under neutral or low pH environment display very little fluctuation. At 350 K there is fluctuation and variation for different pH values, but we cannot find their real difference between (i) their backbone atom RMSDs (root mean square deviations) (Fig. \ref{RMSD_Radius}), which were calculated respectively from the minimized structure, (ii) their radii of gyrations (Fig. \ref{RMSD_Radius}), and (iii) their secondary structures (Fig.s \ref{secondary_structures1}-\ref{secondary_structures2}).\\

Why does HoPrP$^{\text{C}}$ still have stable molecular structures even under low pH environment? Under low pH environment, horse prion protein has 45 and 38  strong hydrogen bonds respectively for {\it seed1} and {\it seed2}, if only considering the resident of hydrogen bonds (HBs) for more than 5\% of the whole 30 ns simulations. All these hydrogen bonds are listed in Table 1 of the Supplementary Materials (where the number before @ is the residue number counted from 1 to 113, i.e. corresponding to residues 119 to 231). By analysis, during the long simulations of 30 ns we can see that all these hydrogen bonds well maintain the three $\alpha$-helices and the two $\beta$-sheets, and especially their interactions linked by the following hydrogen bonds (the first percentage is for {\it seed1} and the second percentage is for {\it seed2}):\\ 
$\bullet$ SER170-TYR218 (linking the $\beta$2-to-$\alpha$2 loop with $\alpha$2, 78.00\%, 65.52\%),\\
$\bullet$ ASP202-TYR157 (linking $\alpha$3 with $\alpha$2, 66.05\%, 65.44\%),\\ 
$\bullet$ TYR157-ARG136 (linking the $\beta$1-to-$\alpha$1 loop with $\alpha$2, 68.27\%, 71.04\%),\\ 
$\bullet$ MET134-ASN159 (linking the $\beta$1-to-$\alpha$1 loop with the $\alpha$1-to-$\beta$2 loop, 29.83\%, 21.47\%),\\
$\bullet$ GLY131-GLN160 (linking the $\beta$1-to-$\alpha$1 loop with the $\alpha$1-to-$\beta$2 loop, 26.67\%, 37.03\%),\\
$\bullet$ HIS140-ARG208 (linking the $\beta$1-to-$\alpha$1 loop with $\alpha$3, 43.85\%, 10.80\%),\\
$\bullet$ SER132-GLN217 (linking the $\beta$1-to-$\alpha$1 loop with $\alpha$3, 29.70\%, 12.90\%),\\
$\bullet$ PRO137-TYR149 (linking the $\beta$1-to-$\alpha$1 loop with $\alpha$1, 17.18\%, 23.39\%),\\
$\bullet$ PRO158-ARG136 (linking the $\beta$1-to-$\alpha$1 loop with $\alpha$2, 9.53\%, 6.01\%),\\
$\bullet$ ARG156-HIS187 (linking $\alpha$1 and $\alpha$2, 6.64\%, 33.70\%), and\\
$\bullet$ GLU221-TYR163 (linking $\alpha$3 and $\beta$2, 11.40\%, 6.22\%).\\
Hydrophobic interactions also contribute greatly to the structural stability of horse prion protein under low pH environment. In all there are 1760 and 2006 hydrophobic bonds respectively for {\it seed1} and {\it seed2} during the whole 30 ns of simulations, and for {\it seed1} and {\it seed2} respectively, there are 383 and 363 strong hydrophobic interactions are 100\% resident in the core of the protein (Table 2 of the Supplementary Materials). The three hydrophobic bonds CYS214-CYS179, CYS214-VAL176, TYR162-LEU130 should be well noticed, where there is a disulfide bond between CYS179 and CYS214 linking $\alpha$-helices 2 and 3, VAL176 is in $\alpha$-helix 2, and TYR162-LEU130 (but 99.06\% for {\it seed2}) are just respectively in the strands 1 and 2 of the antiparallel $\beta$-sheet (we noticed that there is not a hydrogen bond between TYR162 and LEU130).\\ 

\begin{figure}[h!]
\centerline{
\includegraphics[width=6.0in]{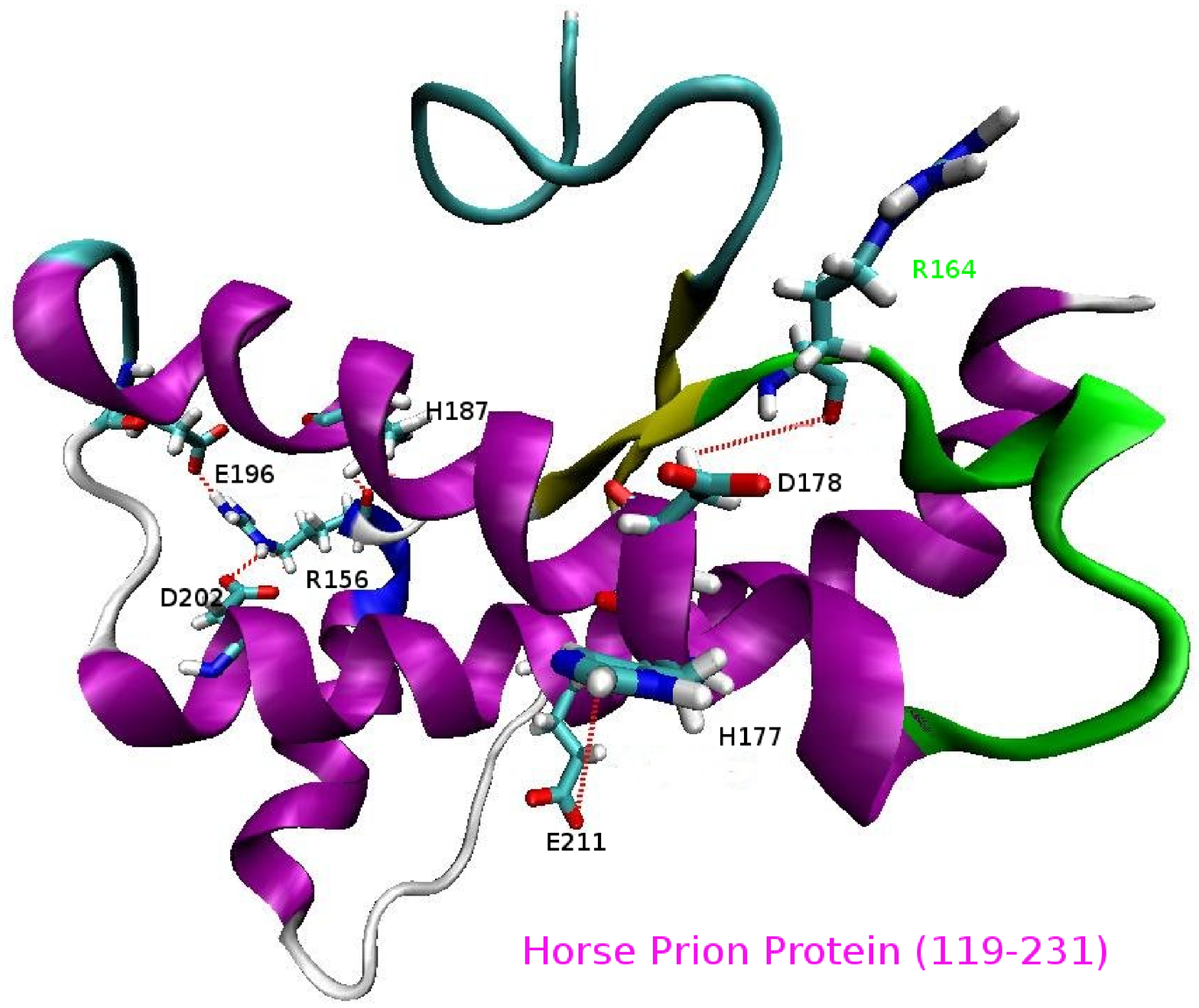}
}
\caption{Salt bridges GLU211-HIS177, GLU196-ARG156-HIS187, ARG156-ASP202 and ASP178-ARG164 of horse prion protein (dash line: a salt bridge between the two residues, green color: the $\beta$2-to-$\alpha$2 loop).}
\label{salt_bridges}
\end{figure}

Under the neutral pH environment, there are 33 salt bridges respectively for {\it seed1} and {\it seed2}, which contribute to the structural stability of horse prion protein (Table 3 of the Supplementary Materials). These salt bridges well keep the structural stability of the three helices and their interactions. During the whole 30 ns, among these salt bridges the following ones are important in the contributions of the three $\alpha$-helices and two $\beta$-sheets' structural stability of horse prion protein: GLU211-HIS177 (occupied rate 27.48\% for {\it seed1} and 17.77\% for {\it seed2}), GLU196-ARG156 (occupied rate 17.17\% for {\it seed1} and 43.47\% for {\it seed2}), ARG156-HIS187 (occupied rate 9.14\% for {\it seed1} and 75.60\% for {\it seed2}), ARG156-ASP202 (occupied rate 0.25\% for {\it seed1} and 6.69\% for {\it seed2}), ASP178-ARG164 (occupied rate 10.45\% for {\it seed1} and 4.30\% for {\it seed2}). The positions of all these important salt bridges can be seen in Fig. \ref{salt_bridges}. We can see that the residue ARG156 of $\alpha$-helix 1 forms a network of three salt bridges (and hydrogen bonds) linking $\alpha$-helices 2 and 3.\\

Recently, some researchers (6, 47, 48, 49, 50, 51, 52) reported that the $\beta$2-$\alpha$2 loop plays an important role to stabilize the structural stability of rabbit and horse prion proteins. This conclusion can be confirmed from the hydrogen bond, hydrophobic bond, and salt bridge databases on horse prion protein presented by this paper. All their reports are not addressing the salt bridge problem of this loop. Like (44-46), this paper also reports the important salt bridge ASP178-ARG164 (like a taut bow string) keeping this loop linked. The important salt bridges reported in this paper, together with the highly ordered $\beta$2-$\alpha$2 loop and its interactions with $\alpha$-helix 3, maintain the structural stability of horse prion protein in a perfect way.\\

\begin{figure}[h!]
\centerline{
\includegraphics[width=5.6in]{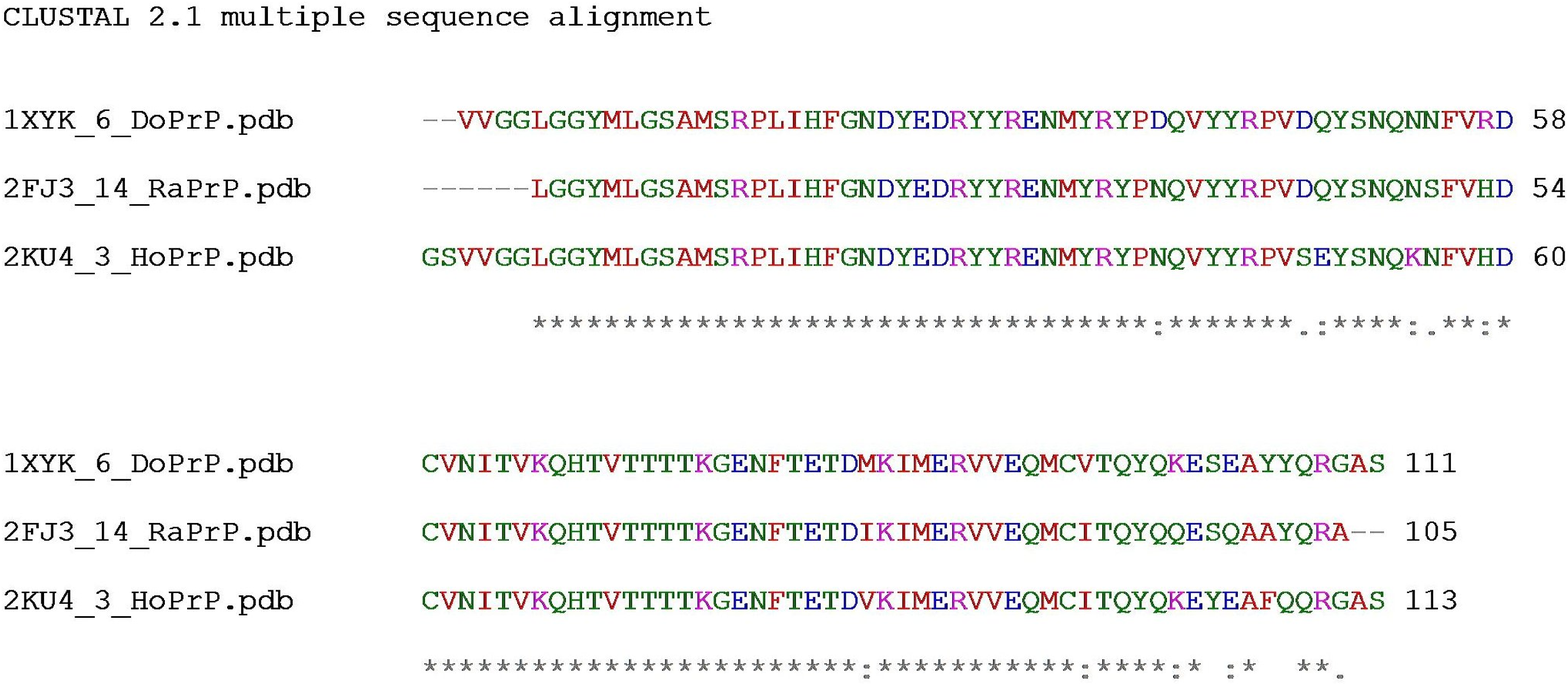}
}
\caption{Multiple sequence alignment of dog, rabbit, and horse prion proteins.}
\label{sequence_alignment}
\end{figure}

Lastly, we make some comparison of horse prion protein with dog and rabbit prion proteins. There sequences alignment is shown in Fig. \ref{sequence_alignment}.
In Fig. \ref{sequence_alignment},  ``*" means that the residues in that column are identical in all sequences in the alignment, ``:" means that conserved substitutions have been observed, ``." means that semi-conserved substitutions are observed, the RED color takes place at small (small+ hydrophobic (incl.aromatic-Y)) residues, the BLUE color takes place at acidic residues, the MAGENTA color takes place at Basic-H residues,GREEN color takes place at Hydroxyl+sulfhydryl+amine+G residues, and Grey color takes place at unusual amino/imino acids etc.. We may see in Fig. \ref{sequence_alignment} that there are more identities and similarities between horse and dog prion proteins, compared with the identities and similarities between horse and rabbit prion proteins, between dog and rabbit prion proteins. This point can be furthermore confirmed by the following data of the Needleman-Wunsch or Smith-Waterman Pairwise Sequence Alignment and the jCE algorithm or jCE Circular Permutation Pairwise Structure Alignment. For horses and dogs, the Identities are 89.38\% (query) and 90.99\% (subject) and the Similars are 93.81\% (query) and 95.50\% (subject) for the Sequence Alignment, and the Identity is 89.11\% and the Similarity is 93.07\% for the Structure Alignment. For horses and rabbits, the Identities are 87.61\% (query) and 71.74\% (subject) and the Similars are 93.81\% (query) and 76.81\% (subject) for the Sequence Alignment, and the Identity is 71.72\% and the Similarity is 77.78\% for the Structure Alignment. For dogs and rabbits, the Identities are 72.46\% (query) and 90.09\% (subject) and the Similars are 76.81\% (query) and 95.50\% (subject) for the Sequence Alignment, and the Identity is 92.93\% and the Similarity is 98.99\% for the Structure Alignment. The 3-Dimensional structural Identity and Similarity of horse and dog prion proteins (89.11\%, 93.07\%) are clearly very larger than those of horse and rabbit prion proteins (71.72\%, 77.81\%). This explains the reasons why rabbit prion protein differs very much from dog and horse prion proteins in secondary structures under low pH environment (45). The rabbit / horse and dog prion proteins have a strong salt bridge ASP177-ARG163 / ASP178-ARG164 (like a taut bow string) keeping the $\beta$2-$\alpha$2 loop linked (44, 45). But the salt bridge ASP178-ARG164 does not exist for human and mouse prion proteins (46). For horses and humans (1QLX.pdb), the Identities are 86.73\% (query) and 46.67\% (subject) and the Similars are 51.43\% (subject) for the Sequence Alignment. For horses and mouses (1AG2.pdb), the Identities are 80.53\% (query) and the Similars are 87.61\% (query) for the Sequence Alignment. Thus, compared with the sequence alignments of horses-dogs (89.38\%, 90.99\%, 93.81\%, 95.50\%) and horses-rabbits (87.61\%, 71.74\%, 93.81\%, 76.81\%), the percentages of Identities and Similars of the sequence alignments of horses-humans (86.73\%, 46.67\%, 51.4\%) and horses-mouses (80.53\%, 87.61\%) are clearly very less. This is due to human and mouse prion proteins are non-resistive prion proteins, but dog, horse and rabbit prion proteins are prion disease resistive proteins. Rabbit prion protein differs from dog prion protein in that, except two residues at positions 158 and 173 are different, all other different residues are in helix 3 (46). However, compared with horse prion protein, rabbit prion protein has different residues at positions 158, 159, 172, and 173. In the C-terminal end of helix 3, at positions 219, 222 and 228 rabbit prion protein differs from dog and horse prion proteins. We also find that horse prion protein has different residues from dog and rabbit prion proteins at positions 167, 168, 173, 212, 216. The residues at all these special positions should specially contribute to the structural stability of these resistive prion proteins.\\

\section*{Conclusion}
Rabbits, dogs, and horses are the only mammalian species reported to be resistant to infection from prion diseases isolated from other species. Recently, the $\beta$2-$\alpha$2 loop has been reported to contribute to their protein structural stabilities. The author has found that rabbit prion protein has a strong salt bridge ASP177-ARG163 (like a taut bow string) keeping this loop linked. This paper confirms that this salt bridge also contributes to the structural stability of horse prion protein. Thus, the region of $\beta$2-$\alpha$2 loop should be a potential drug target region. Besides this very important salt bridge, other four important salt bridges GLU196-ARG156-HIS187, ARG156-ASP202 and GLU211-HIS177 are also found to greatly contribute to the structural stability of horse prion protein. Rich databases of salt bridges, hydrogen bonds and hydrophobic contacts for horse prion protein can be found in the Supplementary Materials of this paper.\\
 
\section*{Supplementary Materials: see the ``J BioMol Strut Dyn" journal website}
\subsection*{I. Videos 1-2}
In Fig. \ref{salt_bridges}, the important salt bridges of initial structure of HoPrP$^{\text{C}}$ (i.e. 2KU4.pdb) are illuminated. Videos 1-2 illuminate the structural dynamics of all these important salt bridges during the whole 30 ns' simulations. Video 1 is for {\it seed1} and Video 2 is for {\it seed2}. For each video, there are 150 frames and there are 5 frames (at the snapshots of 200 ps, 400 ps, 600 ps, 800 ps, 1000 ps) taken for each nanosecond.
\subsection*{II. Tables 1-3}
{\bf Table 1} is the database of the Hydrogen Bonds of horse prion protein under low pH environment at 350 K for both {\it seed1} and {\it seed2}. The detailed descriptions for Tables 1-3 can be found at the header of each database.\\ 
{\bf Table 2} is the database of the Hydrophobic Bonds of horse prion protein under low pH environment at 350 K for both {\it seed1} and {\it seed2}.\\ 
{\bf Table 3} is the database of Salt Bridges of horse prion protein at 300K for both {\it seed1} and {\it seed2}, under neutral pH environment. These three rich databases of Salt Bridges, Hydrophobic and Hydrogen Bonds of horse prion protein should present very useful structural bioinformatics to study horse prion protein and prion diseases.\\

\noindent {\bf {\large Acknowledgments:}} This research was supported by a Victorian Life Sciences Computation Initiative (VLSCI) grant number VR0063 on its Peak Computing Facility at the University of Melbourne, an initiative of the Victorian Government. The author especially appreciates Mr. David Bannon of VLSCI for his great helps and supports to Project VR0063.

\section*{References}
\noindent 1. A. Aguzzi and M.. Heikenwalder. {\it Nat Rev Microbiol} 4, 765-775 (2006).\\

\noindent 2. S. B. Prusiner. {\it Proc Natl Acad Sci USA} 95, 13363-13383 (1998).\\

\noindent 3. C. Weissmann. {\it Microbiol} 2, 861-871 (2004).\\

\noindent 4. M. Polymenidoua, H. Trusheimb, L. Stallmacha, R. Moosa, J. A. Julius, G. Mielea, C. Lenzbauerb, and A. Aguzzia. {\it Vaccine} 26, 2601-2614 (2008).\\

\noindent 5. I. Vorberg, H. G. Martin, P. Eberhard, and A. P. Suzette. {\it J Virol} 77, 2003-2009 (2003).\\

\noindent 6. D. R. Perez, F. F. Damberger, and K. Wuthrich. {\it J Mol Biol} 400(2), 121-128 (2010).\\ 

\noindent 7. J. Yoon, S. Jang, K. Lee, and S. Shin. {\it J Biomol Struct Dyn} 27, 259-269 (2009).\\

\noindent 8. D. Ilda, C. Giovanni, and D. Alessandro. {\it J Biomol Struct Dyn} 27, 307-317 (2009).\\

\noindent 9. A. Cordomi and J. J Perez. {\it J Biomol Struct Dyn} 27, 127-147 (2009).\\

\noindent 10. L. K. Chang, J. H. Zhao, H. L. Liu, K. T. Liu, J. T. Chen, W. B. Tsai, and Y. Ho. {\it J Biomol Struct Dyn} 26, 731-740 (2009).\\

\noindent 11. J. H. Zhao, H. L. Liu, Y. F. Liu, H.Y. Lin, H. W. Fang, Y. Ho, and W. B Tsai. {\it J Biomol Struct Dyn} 26, 481-490 (2009).\\

\noindent 12. P. S. Fang, J. H. Zhao, H. L. Liu, K. T. Liu, J. T. Chen, H. Y. Lin, C. H. Huang, and H. W. Fang. {\it J Biomol Struct Dyn} 26, 549-559 (2009).\\

\noindent 13. P. N. Sunilkumar, D. G Nair, C. Sadasivan, and M. Haridas. {\it J Biomol Struct Dyn} 26, 491-496 (2009).\\

\noindent 14. H. R. Bairagya, B. P. Mukhopadhyay, and K. Sekar. {\it J Biomol Struct Dyn} 27, 149-158 (2009).\\

\noindent 15. M. S. Achary and H. A. Nagarajaram. {\it J Biomol Struct Dyn} 26, 609-623 (2009).\\

\noindent 16. G. Patargias, H. Martay, and W. B. Fischer. {\it J Biomol Struct Dyn} 27, 1-12 (2009).\\

\noindent 17. F. Mehrnejad and M. Zarei. {\it J Biomol Struct Dyn} 27, 551-559 (2010).\\

\noindent 18. A. Sharadadevi and R. Nagaraj. {\it J Biomol Struct Dyn} 27, 541-550 (2010).\\

\noindent 19. I. Tuszynska and J. M. Bujnicki. J Biomol Struct Dyn 27, 511-520 (2010).\\

\noindent 20. P. Sklenovsky and M. Otyepka. J Biomol Struct Dyn 27, 521-539 (2010).\\

\noindent 21. B. Jin, H. M. Lee, and S. K. Kim. {\it J Biomol Struct Dyn} 27, 457-464 (2010).\\

\noindent 22. S. Roy and A. R. Thakur. {\it J Biomol Struct Dyn} 27, 443-455 (2010).\\

\noindent 23. C. Carra and F. A. Cucinotta. {\it J Biomol Struct Dyn} 27, 407-427 (2010).\\

\noindent 24. Y. Yu, Y. Wang, J. He, Y. Liu, H. Li, H. Zhang, and Y. Song. {\it J Biomol Struct Dyn} 27, 641-649 (2010).\\

\noindent 25. Z. Cao and J. Wang. {\it J Biomol Struct Dyn} 27, 651-661 (2010).\\

\noindent 26. S. Sharma, U. B. Sonavane, and R. R. Joshi. {\it J Biomol Struct Dyn} 27, 663-676 (2010).\\

\noindent 27. M. J. Aman, H. Karauzum, M. G. Bowden and T. L. Nguyen. {\it J Biomol Struct Dyn} 28, 1-12 (2010).\\

\noindent 28. Z. X. Cao, L. Liu, and J. H. Wang. {\it J Biomol Struct Dyn} 28, 343-353 (2010).\\

\noindent 29. L. K. Chang, J. H. Zhao, H. L. Liu, J. W. Wu, C. K. Chuang, K. T. Liu, J. T. Chen, W. B. Tsai, and Y. Ho. {\it J Biomol Struct Dyn} 28, 39-50 (2010).\\

\noindent 30. Z. Gong, Y. J. Zhao, and Y. Xiao. {\it J Biomol Struct Dyn} 28, 431-441 (2010).\\

\noindent 31. C. Koshy, M. Parthiban, and R. Sowdhamini. {\it J Biomol Struct Dyn} 28, 71-83 (2010).\\

\noindent 32. H. M. Lee, B. A. Jin, S. W. Han, and S. K. Kim. {\it J Biomol Struct Dyn} 28, 421-430 (2010).\\

\noindent 33. R. Nasiri, H. Bahrami, M. Zahedi, A. A. Moosavi-Movahedi, and N. Sattarahmady. {\it J Biomol Struct Dyn} 28, 211-226 (2010).\\

\noindent 34. Y. Tao, Z. H. Rao, and S. Q. Liu. {\it J Biomol Struct Dyn} 28, 143-157 (2010).\\

\noindent 35. J. F. Varughese, J. M. Chalovich, and Y. M. Li. {\it J Biomol Struct Dyn} 28, 159-173 (2010).\\

\noindent 36. J. Wiesner, Z. Kriz, K. Kuca, D. Jun, and J. Koca. {\it J Biomol Struct Dyn} 28, 393-403 (2010).\\

\noindent 37. Z. W. Yang, N. Wu, Y. J. Fu, G. Yang, W. Wang, Y. G. Zu, and T. Efferth. {\it J Biomol Struct Dyn} 28, 323-330 (2010).\\

\noindent 38. Y. Yuan, M. H. Knaggs, L. B. Poole, J. S. Fetrow, and F. R. Salsbury. {\it J Biomol Struct Dyn} 28, 51-70 (2010).\\

\noindent 39. L. H. Zhong. {\it J Biomol Struct Dyn} 28, 355-361 (2010).\\

\noindent 40. L. Zhong and J. Xie. {\it J Biomol Struct Dyn} 26, 525-533 (2009).\\

\noindent 41. C. Mangels, R. Kellner, J. Einsiedel, P. R. Weiglmeier, P. Rosch, P. Gmeiner, S. Schwarzinger. {\it J Biomol Struct Dyn} 27, 13-22 (2010).\\

\noindent 42. C. Mangels, A. O. Frank, J. Ziegler, R. Klingenstein, K. Schweimer, D. Willbold, C. Korth, P. Rosch, S. Schwarzinger. {\it J Biomol Struct Dyn} 27, 163-170 (2009).\\

\noindent 43. J. P. Zhang. {\it J Biomol Struct Dyn} 27, 159-162 (2009).\\

\noindent 44. J. P. Zhang. {\it J Theor Biol} 264, 119-122 (2010).\\

\noindent 45. J. P. Zhang and D. D. W. Liu. {\it J Biomol Struct Dyn} 28 (6), 861-869 (2011).\\

\noindent 46. J. P. Zhang. {\it J Theor Biol} 269 (1), 88-95 (2011).\\

\noindent 47. P. Fernandez-Funez, Y. Zhang, J. Sanchez-Garcia, K. Jensen, W. Q. Zou, and D. E. Rincon-Limas. {\it Commun Integr Biol} 4 (3), 1-5 (2011).\\

\noindent 48. M. Q. Khan, B. Sweeting, V. K. Mulligan, P. E. Arslan, N. R. Cashman, E. F. Pai, and A. Chakrabartty. {\it Proc Natl Acad Sci USA} 107 (46), 19808–19813 (2010).\\

\noindent 49. D. H. Lin, Y. Wen. {\it Scientia Sinica Chimica} 41 (4), 683-698 (2011).\\

\noindent 50. B. Sweeting, E. Brown, A. Chakrabartty, and E. F. Pai. {\it Canadian Light Source} 28 (2009): 
http://www.lightsource.ca/science/pdf/activity\_report\_2009/28.pdf\\
 
\noindent 51. Y. Wen, J. Li, M. Q. Xiong, Y. Peng, W. M. Yao, J. Hong, and D. H. Lin. {\it PLoS One} 5(10): e13273 (2010).\\

\noindent 52. Y. Wen, J. Li, W. M. Yao, M. Q. Xiong, J. Hong, Y. Peng, G. F. Xiao, and D. H. Lin. {\it J Bio Chem} 285 (41), 31682-31693 (2010).\\ 
\end{document}